\documentclass{article}

\usepackage{PRIMEarxiv}
\usepackage{amsmath}
\usepackage[utf8]{inputenc} % allow utf-8 input
\usepackage[T1]{fontenc}    % use 8-bit T1 fonts
\usepackage{hyperref}       % hyperlinks
\usepackage{url}            % simple URL typesetting
\usepackage{booktabs}       % professional-quality tables
\usepackage{amsfonts}       % blackboard math symbols
\usepackage{nicefrac}       % compact symbols for 1/2, etc.
\usepackage{microtype}      % microtypography
\usepackage{lipsum}
\usepackage{fancyhdr}       % header
\usepackage{graphicx}       % graphics
\graphicspath{{media/}}     % organize your images and other figures under media/ folder

\title{Integrating LSTM and BERT for Long-Sequence Data Analysis in Intelligent Tutoring Systems
\thanks{\textit{\underline{Citation}}: 
\textbf{Authors. Title. Pages.... DOI:000000/11111.}} 
}

\author{
  Zhaoxing Li, Sebastian Stein \\
  School of Electronics and Computer Science\\
  University of Southampton \\
  Southampton\\
  \texttt{\{zhaoxing.li, ss2.esc\}@soton.ac.uk} \\
   \And
  Jujie Yang, Jindi Wang\\
  Department of Computer Science\\
  Durham University \\
  Durham\\
  \texttt{\{jujie.yang, jindi.wang\}@soton.ac.uk} \\
  \AND
  Lei Shi\\
  School of Computing \\
  Newcastle University\\
  Newcastle upon Tyne\\
  \texttt{lei.shi@ncl.ac.uk} \\
}

\begin{document}
\maketitle

\begin{abstract}
The field of Knowledge Tracing aims to understand how students learn and master knowledge over time by analyzing their historical behaviour data. To achieve this goal, many researchers have proposed Knowledge Tracing models that use data from Intelligent Tutoring Systems to predict students' subsequent actions. However, with the development of Intelligent Tutoring Systems, large-scale datasets containing long-sequence data began to emerge. Recent deep learning based Knowledge Tracing models face obstacles such as low efficiency, low accuracy, and low interpretability when dealing with large-scale datasets containing long-sequence data. To address these issues and promote the sustainable development of Intelligent Tutoring Systems, we propose a \textbf{L}STM \textbf{B}ERT-based \textbf{K}nowledge \textbf{T}racing model for long sequence data processing, namely \textbf{LBKT}, which uses a BERT-based architecture with a Rasch model-based embeddings block to deal with different difficulty levels information and an LSTM block to process the sequential characteristic in students' actions. LBKT achieves the best performance on most benchmark datasets on the metrics of ACC and AUC. Additionally, an ablation study is conducted to analyse the impact of each component of LBKT's overall performance. Moreover, we used t-SNE as the visualisation tool to demonstrate the model's embedding strategy. The results indicate that LBKT is faster, more interpretable, and has a lower memory cost than the traditional deep learning based Knowledge Tracing methods.
\end{abstract}

% keywords can be removed

\section{Introduction}

Technology Enhanced Learning (TEL) has become increasingly important in providing high-quality education to build a more sustainable world. The recent COVID-19 pandemic has significantly impacted traditional classroom education and sparked online learning, enabling teaching and learning remotely. Meanwhile, the development of online learning systems has made it possible to use Intelligent Tutoring Systems (ITS) to store and analyse a sizable amount of student behaviour data to improve intelligent educational services. As one of the widely applied TEL technologies, Knowledge Tracing (KT) has drawn a lot of attention. KT is the field of modelling students' learning trajectories and predicting their sequential actions based on historical interaction data between students and ITS \cite{abdelrahman2022knowledge}. 
 
With the development of ITS, large-scale datasets such as \emph{EdNet} \cite{choi2020ednet} and \emph{Junyi Academy} \cite{chang2015modeling} began to emerge. In these datasets, long-sequence student interaction data were gathered as an increasing number of students used the ITS for an extended period. The long- and short-sequence data in these datasets are unbalanced, which satisfies the long-tail distribution \cite{liu2021efficient}. For instance, within the EdNet dataset, a substantial amount of student action sequences are included, ranging from the shortest sequence that may comprise just a single action to the longest sequence that encompasses 40,157 actions. Notably, the average action sequence length of the EdNet dataset is 121.5, indicating a moderate length of data sequences overall. However, it is important to note that the distribution of sequence lengths is highly skewed, and this unbalanced distribution has an impact on the overall performance of the KT models \cite{jeon2021last}. Although the quantity of short-sequence data is larger than the long-sequence data, the latter is of more weight than the former in prediction tasks \cite{liu2021efficient}.

In general, KT models could be divided into three categories: probabilistic KT models, logistic KT models, and deep learning based KT methods (DKT) \cite{liu2021survey}. Traditional probabilistic KT models and logistic KT models are forced to confront difficulties such as decreased processing efficiency and increased memory usage as growing amounts of longer sequence data are released. Deep learning based KT models are known to suffer from inefficiencies when processing long-sequence action data problems, including issues related to the accuracy, speed, and memory usage \cite{jeon2021last,liu2021efficient}. Therefore, allowing the processing of very long sequence data is key to achieving high performance for next-generation KT models \cite{jeon2021last}. Moreover, due to the black-box nature of traditional deep learning methods, the current deep learning based KT models also struggle with the lack of interpretability \cite{ghosh2020context}.

To address the above issues, in this paper, we propose LBKT, a novel \textbf{L}STM \textbf{B}ERT \textbf{K}nowledge \textbf{T}racing model, for processing long sequence data. The model combines the strength of the Bidirectional Encoder Representations from Transformers (BERT) model in capturing the relations of complex data \cite{devlin2018bert} with the strength of the LSTM model in handling long sequential data to improve its performance on large-scale datasets containing long-sequence data (here, the long-sequence data indicates a length longer than 400 interactions). Moreover, we utilise a Rasch model-based embedding method to process the difficulty level information in the historical behaviour data of students. The Rasch model is a classic yet powerful model in psychometrics \cite{rasch1993probabilistic}, which could be utilised to construct raw questions and knowledge embeddings for KT tasks \cite{ghosh2020context}. Rasch model based embedding could improve the model's performance and interpretability. The experimental results show that our proposed LBKT outperforms the baseline models in five datasets on metrics ACC and AUC. Moreover, it is faster at processing long-sequence data at two long-sequence datasets we extract from the two large-scale datasets. Furthermore, we use t-SNE as the visualisation tool to demonstrate the interpretability of the embedding strategy.

The main contributions of our paper lie in the following two aspects:

\begin{enumerate}

\item We propose LBKT \footnote{Source code and datasets are available at https://github.com/******/LBKT}, a novel \textbf{L}STM \textbf{B}ERT \textbf{K}nowledge \textbf{T}racing model for long sequence data processing. The LBKT leverages the power of BERT, Rasch-based embedding strategies, and LSTM.  

\item The experimental results show that LBKT outperforms the baseline models on five ITS datasets on the metric of AUC(assist12, assist17, algebra06, EdNet, and Junyi Academy). Another comparative experiments show the effectiveness of LBKT when processing long-sequence datasets. LBKT model exhibits better interpretability than traditional deep learning based KT models and has advantages in training efficiency.
\end{enumerate}

\section{Related Work}

\subsection{Knowledge Tracing}
Knowledge Tracing (KT) is used in Intelligent Tutoring Systems (ITS) to model and predict a student's mastery level of a specific skill or concept over time \cite{abdelrahman2023knowledge}. It is based on the assumption that a student's knowledge state is a hidden variable that can be inferred from their observable behaviour, such as their responses to questions or tasks related to the skill or concept being measured \cite{corbett1994knowledge}. Its goal is to provide personalised feedback and support to students by tracking their progress and adapting instruction to meet their individual needs. This can help to improve student learning outcomes and enhance educational effectiveness. Broadly, there are three categories of KT methods: probabilistic KT models, logistic KT models, and deep learning-based KT models \cite{liu2021survey}.

Probabilistic KT models assume that the student's learning process follows a Markov Process, where students' knowledge mastery could be measured by their observed learning performance \cite{corbett1994knowledge}. Bayesian KT, or BKT, is the earliest and most classic probabilistic model, which was inspired by cognitive mastery learning \cite{corbett2000cognitive}. BKT models generally use a probabilistic graphical model, such as Hidden Markov Model (HMM)\cite{corbett1994knowledge} and Bayesian Belief Network \cite{villano1992probabilistic}, to track students' changing learning states. The major shortcoming of BKT is that it assumes a simplistic two-state student modelling framework, where a student's knowledge is either learned or unlearned, and there is no concept of forgetting or decay in the model. However, in reality, a student's knowledge could be complex and multi-faceted and could change over time due to various factors such as decay and interference. Therefore, BKT may not be able to capture the nuances of student learning and may not provide an accurate representation of their knowledge state over time. For example, BKT assumes that each question only required one skill and that the various skills were irrelevant to each other \cite{corbett1994knowledge,yudelson2013individualized}. Therefore, in general, BKT models cannot process complicated problems, including the multiple skills and the complex relationship among the concepts, questions, and skills. To address this limitation, Käser \textit{et al.} proposed Dynamic BKT, or DBKT, based on Dynamic Bayesian Network (DBN), to model the prerequisite hierarchies and dependencies of multiple skills \cite{kaser2017dynamic}. However, both BKT and DBKT still struggle with processing multiple topics or skills, failing to account for contextual factors that may impact student learning.

The logistic KT models are built on the principle of logistic regression, which is a statistical method used to model the probability of a binary outcome based on one or more predictor variables \cite{liu2021survey}. In the context of educational data, the predictor variables could include a student's prior performance on a set of related skills or concepts, their response time, and their correctness or incorrectness in answering assessment questions. The output of the logistic regression KT model is a probability estimate of a student's mastery level on a particular skill or concept, which can be used to inform personalized learning interventions and improve student outcomes. There are three logistic models. The Learning Factor Analysis model (LFA) incorporates the initial knowledge state, easiness of knowledge components (KCs), and learning rate of KCs to estimate the student's initial knowledge state, the easiness of different KCs, and the learning rate of KCs \cite{cen2006learning}. The Performance Factor Analysis (PFA) model is an extension of the LFA model and takes into account the student's performance. PFA considers parameters for previous failures (f) and successes (s) for the KC, in addition to the easiness of KCs \cite{pavlik2009performance}. The Knowledge Tracing Machines (KTM) model uses factorization machines (FMs) to extend logistic models to higher dimensions \cite{kaser2017dynamic}.

Inspired by the recent success of deep learning (DL)\cite{lecun2015deep}, researchers have applied deep learning technologies into the KT field to develop DL-based Knowledge Tracing \cite{piech2015deep}. DL-based KT typically models a knowledge tracing task as a sequence prediction problem. With the self-attention architectures applied in the deep learning field, KT models based on the self-attention mechanism began to emerge. For example, SAKT \cite{pandey2019self} and SAINT+ \cite{shin2021saint+} apply the self-attention mechanism to KT models and achieve higher performance than the traditional DL-based methods. With the development of the self-attention mechanism, Transformer based knowledge tracing models also have been proposed. Ghosh\cite{ghosh2020context} proposes context-aware attentive knowledge tracing (AKT), which introduces a novel monotonic attention mechanism that accounts for the temporal nature of the learning process and the decay of students' knowledge. Nakagawa \textit{et al.} proposed the Graph-based Knowledge Tracing (GKT) model, which incorporates the potential graph structure of KCs into a graph \cite{nakagawa2019graph}. There were also KT methods based on BERT that had been proposed. MonacoBERT \cite{lee2022monacobert} is a BERT-based KT model that incorporates the monotonic convolutional multi-head attention and classical test-theory-based (CTT-based) embedding strategy to improve performance. BEKT \cite{tiana2021bekt} is a Bidirectional Encoder representation from the Transformers-based model that predicts student knowledge state by combining historical learning performance.

\subsection{Transformer-based Model and Application}

Transformer is a prominent neural network model proposed by Vaswani \textit{et al.}, which utilises the self-attention mechanism to extract inherent features. Transformer-based models have achieved significant success in the Deep Learning field, especially in Nature Language Processing (NLP) and image generation tasks \cite{kalyan2021ammus,parmar2018image}.

The evolution of Transformer-based models, such as BERT \cite{devlin2018bert} and GPT\cite{floridi2020gpt}, has achieved outstanding performance in the above tasks. BERT, first proposed by Devlin \textit{et al.}, is a successful application of Transformer \cite{devlin2018bert}. BERT utilises the self-attention mechanism and the masked language model (MLM) to train the Transformer bidirectionally in the NLP fields \cite{devlin2018bert}. BERT is renowned for its exceptional ability to process and comprehend natural language text efficiently. It has consistently outperformed other deep learning models in a broad range of tasks, extending beyond the field of NLP. BERT's success can be attributed to several key features, including its bidirectional context, which allows it to capture the dependencies between both preceding and succeeding tokens in a sequence. Additionally, BERT's large pre-training corpus enables it to learn a robust language representation that can be fine-tuned for downstream tasks with relatively small amounts of labelled data. BERT's transformer architecture, which uses self-attention mechanisms to capture global dependencies between tokens in a sequence, is also a significant factor contributing to its performance. The self-attention mechanism allows BERT to weigh the importance of different tokens in a sequence dynamically, which improves its ability to capture complex patterns and relationships in the data \cite{devlin2018bert}. BERT is also known for its ability to generate high-quality embeddings, which are crucial for many natural language processing tasks \cite{devlin2018bert}. There have been a lot of BERT variants applied in other deep learning fields, demonstrating their outstanding performances. For example, ConvBERT \cite{jiang2020convbert} applies the original BERT architecture in the image processing field; BERT4Rec uses BERT model to improve recommendation systems \cite{sun2019bert4rec}; LakhNES uses BERT model to enhance Music Generation \cite{donahue2019lakhnes}. However, in the Knowledge Tracing field, although some BERT-based models, such as BEKT \cite{li2023broader,li2022simstu,li2023towards,li2023deep,li2023sim,wang2023exploring,wang2024comparative,wang2023user,wang2024impact} and BiDKT \cite{tan2022bidkt}, are proposed to improve performance, they are unable to outperform state-of-the-art KT methods in large-scale datasets containing long-sequence data.

\section{Methodology}
\subsection{Problem Statement}
The key to knowledge tracing is to predict the correctness of a student's next answer in a sequence. Let $x_1, \ldots, x_t$ denote the student's actions, and let the $t$-th action be represented as $x_t = (q_t, a_t)$, where $q_t$ is the question presented to the student and $a_t$ is the student's response. The goal is to estimate the correctness $P(a_t = 1 | x_1, \ldots, x_{t-1})$, that is, the correctness of student's response to the current question, given their previous actions in the sequence.

\subsection{Proposed Model Architecture}
\begin{figure}
\includegraphics[width=\textwidth]{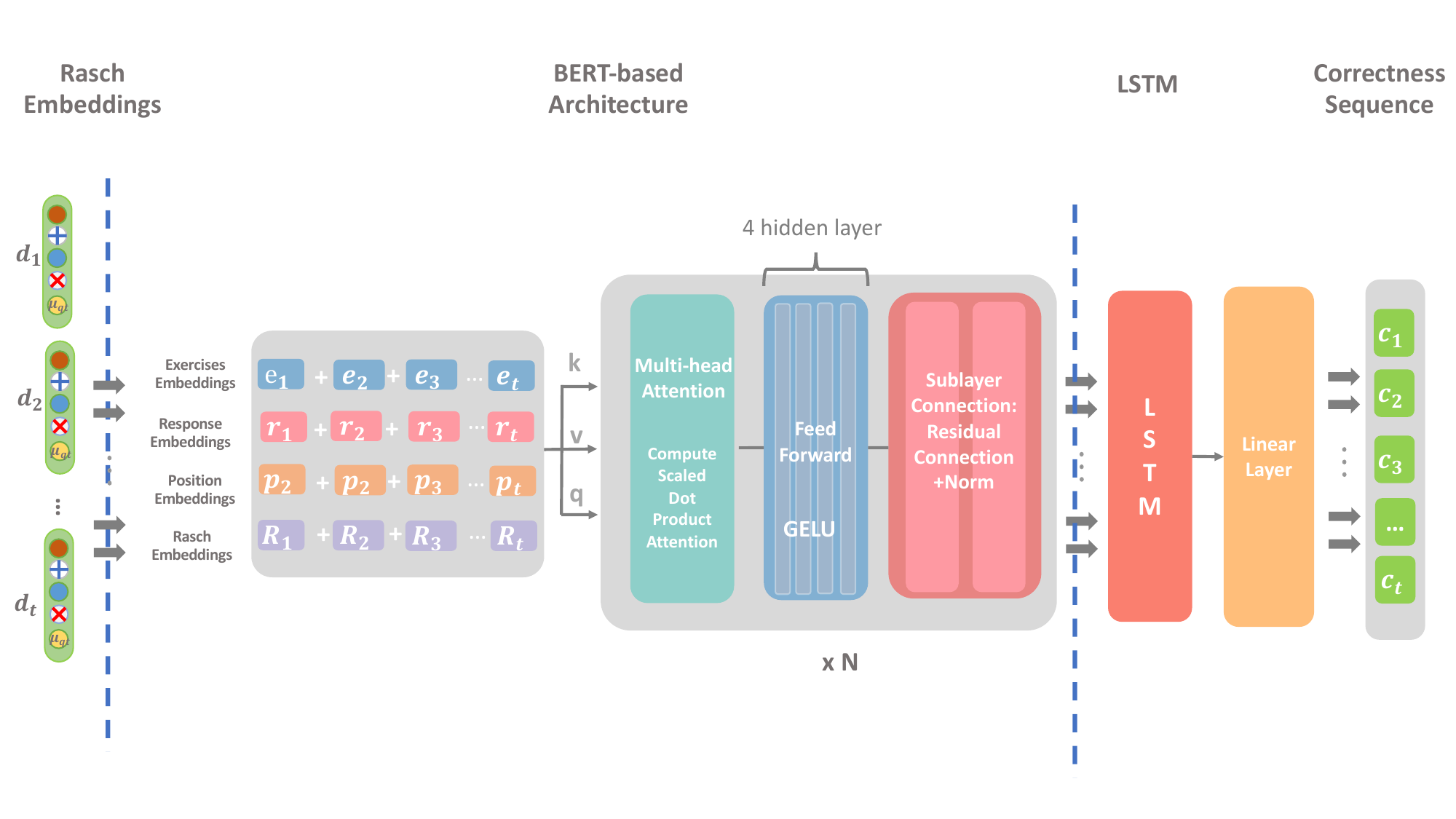}
\centering
\caption{The architecture of LBKT. LBKT consists of three components: 1) the Rasch model-based embeddings (on the left), 2) the BERT-based architecture (in the middle), and 3) the LSTM block (on the right).} \label{architecture}
\end{figure}

We propose a novel model, LBKT, for the task of knowledge tracing on large-scale datasets containing long-sequence data. While previous BERT-based KT models have shown remarkable success in capturing the relations of complex data, they also have inefficiencies when dealing with long sequence student action data \cite{tiana2021bekt}. On the other hand, LSTM models have been proven to excel in handling long sequential data. In response to these challenges, we propose a novel KT model that combines the strengths of both the BERT and LSTM models to improve performance on large-scale datasets containing long-sequence data (where long-sequence data indicates a length longer than 400 interactions). The Rasch embedding (also known as the 1PL IRT model) is a method to represent questions and concepts in a mathematical space \cite{rasch1993probabilistic}. The embeddings are created using a vector that summarizes the variation in questions covering a concept and a scalar difficulty parameter that controls how far a question deviates from the concept it covers. The embeddings are used as raw embeddings for questions and responses, which is a way to track a learner's knowledge state. By leveraging the strengths of a BERT-based model, Rasch model-based embeddings, and long short-term memory (LSTM) unit, our proposed model architecture has the potential to effectively process and understand relationships among different features in long-sequence data, as illustrated in Fig. \ref{architecture}.

The first component of LBKT is the Rasch model-based embeddings proposed by Ghosh \cite{ghosh2020context}. The Rasch model-based embeddings consist of difficulty level embeddings $E_{\text{d}}$ and question embeddings $E_{\text{q}}$. These embeddings are multiplied and added to the BERT token embeddings and the $sin$ and $cos$ positional embeddings to build the final embeddings, as shown in the following equation:

\begin{equation}
    E = E_{\text{Rasch}} + E_\text{BERT Token} + E_\text{Position}
\end{equation}

where the Rasch model-based embeddings $E_{\text{Rasch}}$ are defined as:

\begin{equation}
    E_{\text{Rasch}} = E_{\text{d}} + E_{\text{d}} \times E_{\text{q}}
\end{equation}

The segment embeddings, which are typically used to represent information about the segment in the BERT model, are replaced by the Rasch embeddings mentioned above in our model's architecture. Rasch model-based embeddings are able to more accurately estimate students' knowledge states, as explained earlier, making them a key contributor to the effectiveness of LBKT for knowledge tracing tasks.

The second component of LBKT is a BERT-based block, which consists of 12 Transformer blocks. Each includes a multi-head attention mechanism, a feedforward network (FFN), and sublayer connections. The multi-head attention mechanism uses the ``Scaled Dot Product Attention'' method as implemented in BERT, along with queries $Q$, keys $K$, values $V$, and an attention mask for padded tokens. The FFN has a feedforward hidden layer with a size of four times that of the model's hidden layer and uses the GELU activation function rather than RELU. 

The sublayer connections in the Transformer block include a residual connection followed by layer normalization. The formulas for the attention mechanism and the FFN are as follows:

\begin{equation}
    \operatorname{Attention}(Q, K, V)=\operatorname{softmax}\left(\frac{Q K^T}{\sqrt{d_k}}\right) V
\end{equation}

\begin{equation}
    \text{FFN}(x) = \text{GELU}(W_1x+b_1)W_2+b_2
\end{equation}

In the third component of LBKT, we use a neural network (NN) linear transformation instead of the attention projection typically used in conjunction with the LSTM unit. This is based on our observed improved performance with the NN linear transformation in our experiments. It should be noted that this choice is not necessarily related to the length or complexity of the sequence but rather to the specific characteristics of the data and the task at hand.

Overall, LBKT is a model that is tailored specifically for use in the field of knowledge tracing. It combines the natural language processing capabilities of the BERT model with the ability to accurately estimate knowledge states using Rasch model-based embeddings and the ability to effectively handle long sequences of data using the LSTM unit and the NN linear transformation. This makes it an ideal choice for the task of knowledge tracing in large-scale datasets containing long-sequence data with unbalanced data distribution.

\subsection{Experiment Setting}

\subsubsection{Datasets}

We used five benchmark datasets to validate the effectiveness of the LBKT model, including assist12\footnote{https://sites.google.com/site/assistmentsdata/home}, assist17\footnote{https://sites.google.com/site/assistmentsdata/home}, algebra06 \footnote{https://pslcdatashop.web.cmu.edu/KDDCup}, EdNet \cite{choi2020ednet} \footnote{https://github.com/riiid/ednet}, and Junyi Academy\cite{chang2015modeling}\footnote{https://pslcdatashop.web.cmu.edu/Files?datasetId=1275}. Table \ref{dataset} shows the sizes of the above datasets. In general datasets, such as assist 12 and assist 17, it could be challenging to identify and extract large amounts of long-sequence data. Therefore, we validated the speed performance of every model on two datasets with long-sequence student action data extracted from EdNet and Junyi Academy. The mean action sequence length of EdNet is 121.5. The mean interaction length of Junyi Academic is 104.7. Table \ref{mean} shows the action sequence length statistics of EdNet and Junyi Academy. Here, we define the longer action sequence as longer than 100 records. We extract 200 students' action sequences that include interactions longer than 100 actions from each dataset as the long-sequence dataset to validate the performance of different KT models. Lastly, we selected different lengths of action sequences from Ednet to test the speed performance of each model. We selected four groups with average records lengths of 100, 200, 300, and 400, respectively. Each of these groups included 50 students.

\begin{table}[htbp]
    \caption{Benchmark dataset data statistics}
  \centering
    \setlength{\tabcolsep}{1.7mm}
    \begin{tabular}{ccccccccccc}
\hline
Dataset & Students & Concepts & Questions & Interactions \\
\hline

assist12 & 24,429 & 264 & 51,632 & 1,968,737 \\

assist17 & 1,708 & 411 & 3,162 & 934,638 \\

algebra06 & 1,318 & 1,575 & 549,821 & 1,808,533 \\

EdNet & 784,309 & 1,472 & 11,957 & 641,712 \\

Junyi Academy & 247,606 & 13,169 &  722 &   25,925,922 \\
\hline
\label{dataset}
    \end{tabular}%

\end{table}%

\begin{table}[h]
\centering
\caption{Action Length Statistics of EdNet and Junyi Academy}
\begin{tabular}{ccc}
\hline
   Features & Junyi Academy & EdNet \\
\hline
Students & 247606 & 784309 \\
% Minimum record length & 1 & 1 \\
Max action length & 22067 & 40157 \\
Mean action length & 104.7 & 121.5 \\
\hline
\end{tabular}
\label{mean}

\end{table}
\subsubsection{Baseline Models}
We compared our LBKT to three state-of-the-art models, BEKT \cite{tiana2021bekt}, AKT \cite{ghosh2020context}, DKVMN \cite{sun2019muti}, as well as the two top baseline models in the Riiid Answer Correctness Prediction Competition provided by Kaggle\footnote{https://www.kaggle.com/code/datakite/riiid-answer-correctness}, including SSAKT\cite{zhang2021sequential}, and LTMTI\cite{choi2020ednet}.

\subsubsection{Evaluation Metrics and Validation}

We used the accuracy (ACC) and the area under the curve (AUC) as performance metrics to compare the models' performance in five datasets. We also used the training speed, speed ratio, and memory usage as metrics to compare the performance in the large-scale datasets containing long-sequence data (i.e., EdNet and Junyi Academy). Moreover, we used five-fold cross-validation for the evaluation.

\subsubsection{Hyperparameters for Experiments}
To compare with each model, the same parameters were used for model training. The batch size was set to 64, and the train/test split was 0.8/0.2. The model used an embedding size of 128 and the Adam optimizer with a learning rate of 0.001. The loss function used was the Binary Cross Entropy with Logits Loss (BCEWithLogitsLoss). The scheduler was set to OneCycleLR with a maximum learning rate of 0.002. Dropout was also being used at a rate of 0.2. The training ran for a total of 100 epochs, with early stopping set to 10 epochs. If the validation loss does not decrease for the first three epochs, the training stops, in order to prevent overfitting and save resources. The maximum sequence length was 200, with an eight-attention head. Hidden sizes were 128 for BERT, 512 for FFN, and 128 for LSTM. The Transformer block/encoder layer was set to 12.

\section{Results and Discussion}

\subsection{Overall Performance}
LBKT outperforms four baseline models on most metrics in the experiments on five benchmark datasets. Tabel \ref{tab1} shows the overall performance of each model. We used five-fold cross-validation to estimate their performances. LBKT performed the best on EdNet and Junyi Academy datasets on both ACC and AUC metrics. It also achieved the best performance on the ACC metric on assist12 and AUC on assist17. On algebra06, AKT achieved the best performance on the ACC metric, BEKT achieved the best performance on the AUC metric, and LBKT achieved the second-best performance on both metrics. This result indicates that LBKT is an efficient KT model on most datasets, especially large-scale datasets containing long-sequence interaction data. This was affected by our LBKT model's unique architecture. The LSTM block enables the model to learn the sequential features of the long sequence and gives more importance to the recent actions of the students, which prevents the model from giving too much weight to the long-ago and low-relevance actions and thus improving the training efficiency.

\begin{table}[htbp]
    \caption{Comparison of different KT models on five benchmark datasets. The best performance is denoted in bold.}
  \centering
    \setlength{\tabcolsep}{1.7mm}
    \begin{tabular}{ccccccccccc}
\hline

    Dataset & Metrics & LBKT & BEKT & SSAKT  & LTMTI & AKT & DKVMN  \\
\hline
    assist12 & ACC  & \textbf{0.815} &  0.786  & 0.675  & 0.813  & 0.769 &  0.756\\
     & AUC & 0.768  & \textbf{0.812}  & 0.741  & 0.785  & 0.753   & 0.701 \\
\hline
    assist17 & ACC  &0.792\ & 0.795  & 0.771  & 0.796 & 0.733  &\textbf{ 0.797} \\
     & AUC  & \textbf{0.814}  & 0.801  & 0.735  & 0.683  & 0.803  & 0.709\\
\hline
    algebra06 & ACC  & 0.801  &  0.797 & 0.795  & 0.811  & \textbf{0.831} & 0.800\\
    & AUC & 0.799  &  \textbf{0.815}  & 0.774 & 0.791  & 0.814 & 0.793\\
\hline
    EdNet & ACC  & \textbf{0.803}  & 0.781 & 0.761  & 0.799 &0.756& 0.800 \\
     & AUC & \textbf{0.815} &  0.795  & 0.798 & 0.802  & 0.798   & 0.796\\
\hline
    Junyi  & ACC & \textbf{0.832} &  0.807  & 0.777  & 0.797  & 0.791 & 0.790\\
   Academy  & AUC & \textbf{0.851} & 0.831  & 0.845  & 0.812 & 0.799 & 0.769 \\
\hline

    \end{tabular}%
     
  \label{tab1}%
\end{table}%

Tabel \ref{tabSpeed1} shows the performance comparison on the two large-scale datasets. On both datasets, LBKT achieved the best training efficiency. It was 4.29x faster than BEKT on EdNet and 4.77x faster than BEKT on Junyi Academy. Compared with the second-best model, AKT, LBKT was 1.32x faster on EdNet and 1.42x faster on Junyi Academy. For the memory cost, LBKT was about one-third of BEKT and lower than LTMTL on both datasets. Although the memory cost of LBKT was not the smallest, LBKT has achieved the best results in both ACC and AUC metrics running on the same GPU. This allows LBKT to run on middle-range GPUs.
To improve the training efficiency, we used a last input as the query method \cite{jeon2021last} in the Transformer block instead of the whole sequence, which decreased the complexity of the encoder to improve training speed and reduce memory cost.

\begin{table}[htbp]
    \caption{Performance comparison on the two large-scale datasets, EdNet and Junyi Academy. The best performance is denoted in bold.}
  \centering
    \setlength{\tabcolsep}{1.7mm}
    \begin{tabular}{c|ccc|ccc}
\hline

    Model &  & EdNet &  &   & Junyi Academy & \\
\hline
   &speed $\uparrow$  &speed ratio $\uparrow$  & memory $\downarrow$ & speed$\uparrow$  &speed ratio $\uparrow$  & memory $\downarrow$\\
\hline
    BEKT & 4.93  & 1.00x  & 16.7 GB & 4.85  & 1.00x & 16.6 GB \\
    SSAKT & 7.13 &1.44x & \textbf{3.4 GB} & 6.22 & 1.28x & \textbf{3.2 GB} \\
    LTMTI & 13.8 &1.32x  & 7.69 GB  &12.1 &1.19x & 8.82 GB \\
    AKT & 17.1 & 3.25x & 4.32 GB & 16.4 & 3.35x & 4.37 GB \\
    DKNMN & 5.97 & 2.34x & 7.68 GB & 4.67 & 3.75x & 8.53 GB  \\
    LBKT & \textbf{21.3} & \textbf{4.29x}& 6.09 GB & \textbf{22.2} & \textbf{4.77x} & 6.08 GB \\
   
\hline
    \end{tabular}%
  \label{tabSpeed}%
\end{table}%

To estimate the model performance on different lengths of data sequences, we sorted the data in the two datasets by length and divided them into four sub-datasets according to the average length. The average lengths of the four sub-datasets were set as 100, 200, 300, and 400, respectively. Sequences shorter than the mean were padded with 0s, and sequences longer than the mean were pruned. 

\begin{figure}
\includegraphics[width=\textwidth]{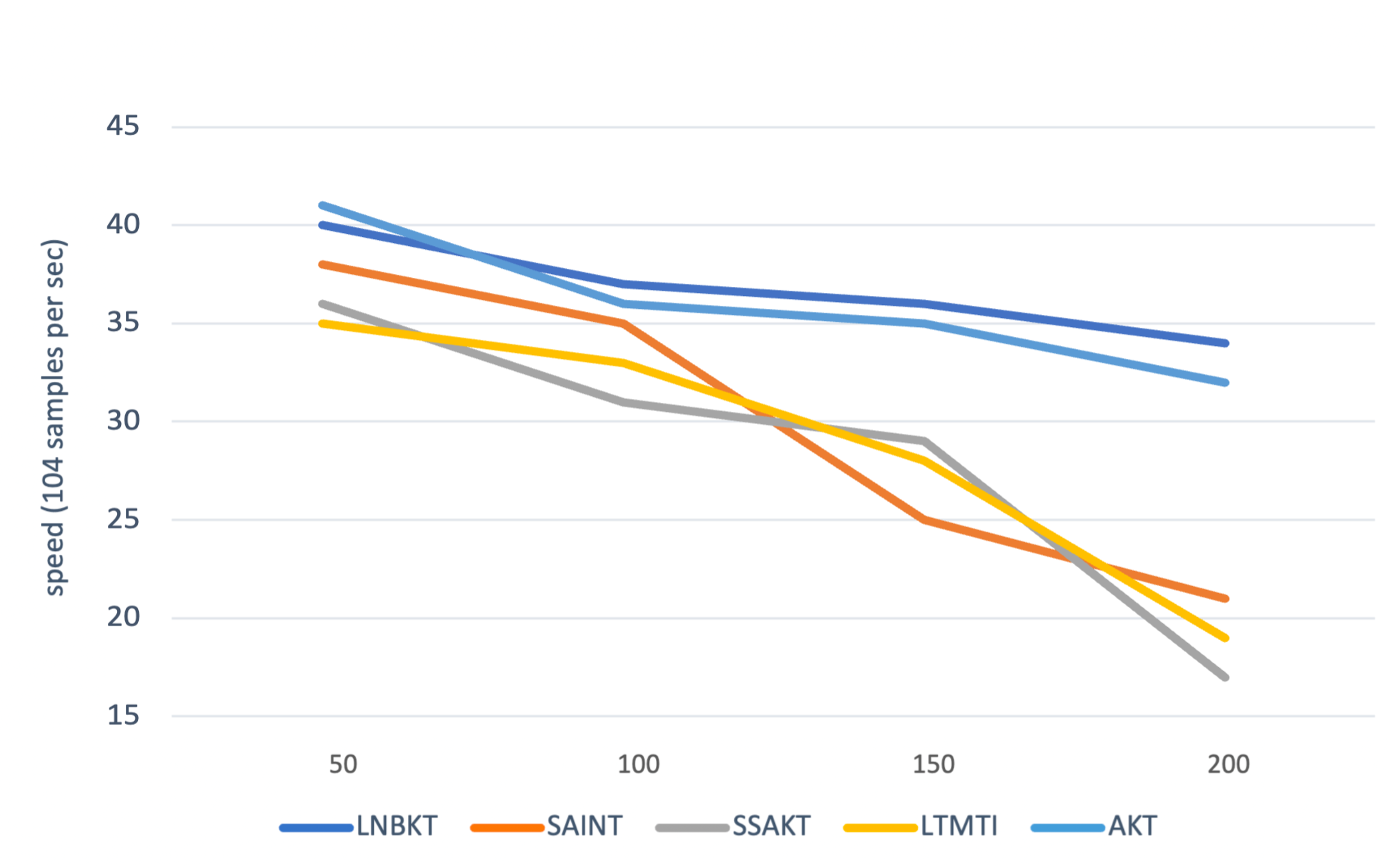}
\centering
\caption{Speed performance comparison of each model when processing data sequences with varying lengths. The vertical axis is the speed ($10^4$ samples per sec).} \label{Speed}
\end{figure}

Fig. \ref{Speed} shows the results of the speed performance comparison of each model processing different lengths of data sequences. LBKT has a relatively high-speed performance compared to other KT models when processing data sequences with varying lengths. LBKT is the fastest model in all four groups of data lengths. AKT and DKMN also have relatively high speeds, with AKT being the second-fastest model in all groups and DKMN being the third-fastest model. Overall, the results suggest that LBKT is the fastest model, and that it is particularly efficient at dealing with long sequences of data. The fact that LBKT maintains its high speed even when processing longer sequences of data indicates that it is well-suited for tasks that require the analysis of large amounts of data over extended periods of time.

\begin{table}[htbp]
    \caption{Results of the ablation study. LBKT-Rasch denotes LBKT without Rasch embedding; LBKT-LTSM denotes LBKT without LSTM block; and BERT denotes only the transformer structure-based blocks are included. The best performance is denoted in bold.}
  \centering
    \setlength{\tabcolsep}{1.7mm}
    \begin{tabular}{ccccccccccc}
\hline

    Dataset & Metrics & LBKT & LBKT-Rasch & LBKT-LSTM  & BERT \\
\hline
    assist12 & ACC  & \textbf{0.804} & 0.785  &  0.799   & 0.793 \\
     & AUC & 0.768 & 0.768 & \textbf{0.783}&  0.750  \\
\hline
    assist17 & ACC  & 0.784  & \textbf{0.792}  & 0.782  & \textbf{0.792}  \\
     & AUC  & \textbf{0.814}  & 0.709  &0.779 & 0.799\\
\hline
    algebra06 & ACC  & \textbf{0.801}& 0.796& 0.792 & 0.798 \\
     & AUC   & 0.799  & 0.756  & \textbf{0.809} & 0.765 \\
\hline
    EdNet & ACC  & \textbf{0.803} &0.729& 0.722 & 0.801  \\
     & AUC  & \textbf{0.815 } & 0.758  & 0.794  &  0.809 \\
\hline
    Junyi Academy & ACC & \textbf{0.882}  & 0.856  & 0.874  & 0.879 \\
     & AUC & \textbf{0.907} & 0.893   & 0.877  & 0.901 \\
\hline

    \end{tabular}%
  \label{tabAblation}%
\end{table}%

\subsection{Ablation Study}

In this section, we explore why LBKT performed better than other methods and which components affected the overall performance. Table \ref{tabAblation} shows the results of the ablation study. We compared LBKT, LBKT without Rasch model-based embeddings block (denoted as LBKT-Rasch), LBKT without LSTM block (denoted as LBKT-LSTM), and LBKT without both Rasch model-based embeddings LSTM (denoted as BERT). The results show that LBKT achieved the best performance on EdNet and Junyi Academy on both ACC and AUC metrics. It also achieved the best performance on one metric in every dataset. BERT-only achieved the best performance on assist17 on ACC, which shows that the combination with Rasch embeddings and LSTM could improve the performance of a single BERT model.

Table \ref{tabSpeed} shows the ablation study of speed performance comparison on the two large-scale datasets. The results show that LBKT has the highest speed performance among all models on both datasets, with a speed of 21.3 samples per second on EdNet and 22.2 samples per second on Junyi Academy. This suggests that LBKT is a highly efficient model for processing large amounts of data in real-time. Interestingly, LBKT-LSTM, which removes the LSTM layer from the proposed model, has a significantly lower speed performance compared to LBKT, with a speed ratio of only 1.29x on EdNet and 1.19x on Junyi Academy. This suggests that the LSTM layer is an important component in the proposed model and contributes significantly to its speed performance. This is likely due to the ability of LSTM to capture long-term dependencies and sequential patterns in the data, which can be crucial in educational applications.

\begin{table}[htbp]
    \caption{Speed performance comparison ablation study on the two large-scale datasets, EdNet and Junyi Academy. The best performance is denoted in bold.}
  \centering
    \setlength{\tabcolsep}{1.7mm}
    \begin{tabular}{c|ccc|ccc}
\hline

    Model &  & EdNet &  &   & Junyi Academy & \\
\hline
    &speed &speed ratio & memory& speed &speed ratio & memory\\
\hline
  LBKT & \textbf{21.3} & \textbf{4.29x}& 6.09GB & \textbf{22.2} & \textbf{4.77x} & 6.08GB \\ 
  
    LBKT-Rash & 19.1 &3.51x & 5.6GB & 20.31 & 3.67x & \textbf{3.1 GB} \\
    LBKT-LSTM & 13.8 &1.29x  & 7.87GB  &12.1 &1.09x & 8.37GB \\
    BERT & 12.1 & 3.27x & \textbf{4.32GB} & 11.4 & 3.52x & 4.34GB \\
    
\hline
    \end{tabular}%
  \label{tabSpeed1}%
\end{table}%

\subsection{Analysis of Embedding Strategy}
\begin{figure}
    \includegraphics[width=\textwidth]{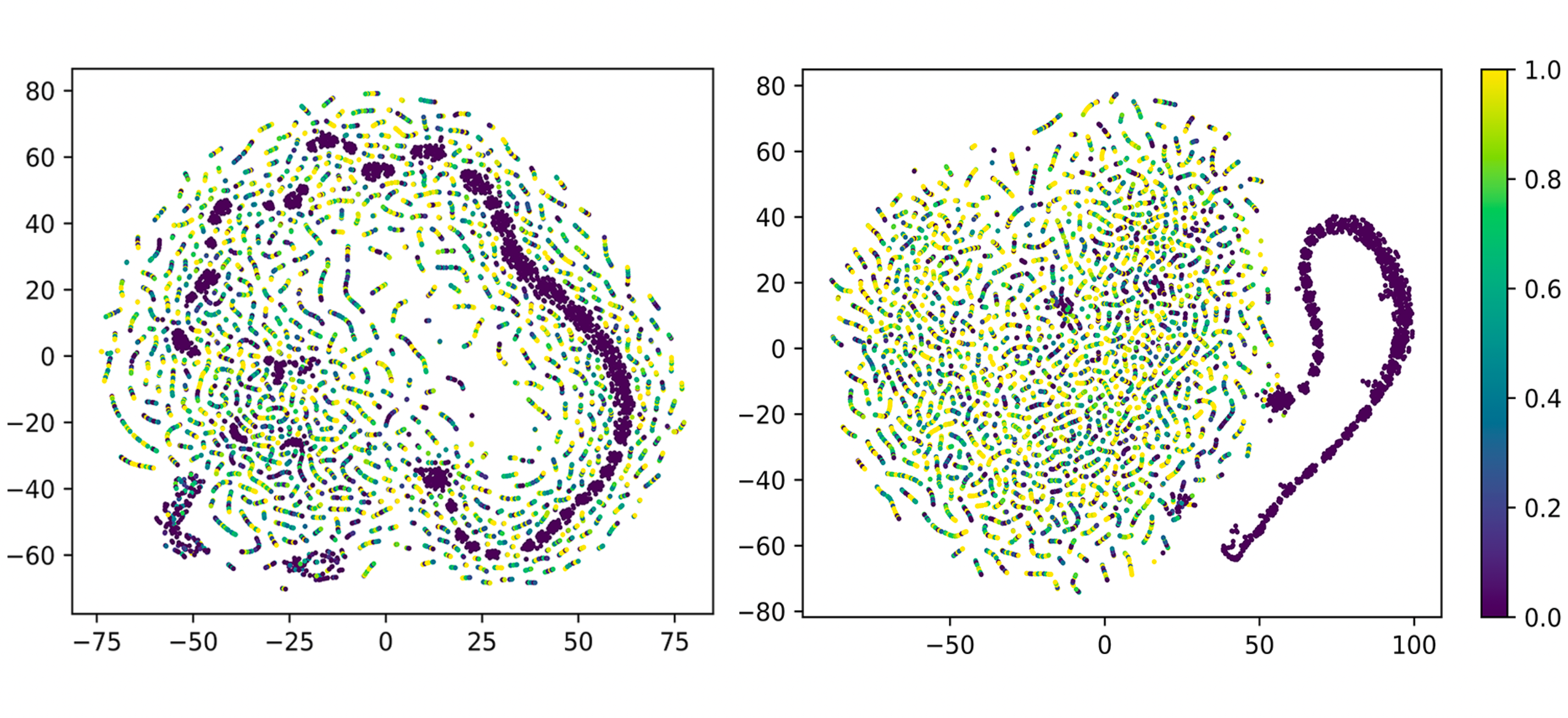}
    \centering
    \caption{Visualisation of the embedding vector using t-SNE: \textit{without} Rasch embeddings (on the left) and \textit{with} Rasch embeddings (on the right). The colour bar is the predicted probability of the outputs.} \label{Vis}
\end{figure}

In this section, We used t-SNE as the visualisation tool to show the interpretability of LBKT's embedding strategy. Fig. \ref{Vis}-\textit{left} shows the results of No-Rasch-embedding, and Fig.\ref{Vis}-\textit{right} shows the Rasch embedding strategy. We can see that, in the No-Rasch-embedding scenario, the difficult questions' embeddings (dark blue vectors) mixed with the easy questions' embeddings (yellow to light blue vectors). In figure \ref{Vis}-\textit{right}, the difficult level embeddings were separated to avoid mixing with easy level embeddings.

Questions at a higher difficulty level are typically associated with longer sequence data, as students spend more time and steps on difficult exercises, which results in longer interaction sequences. Rasch model-based embeddings could divide different difficulty-level parts before the start of the model training and not mix them with other difficulty-level embeddings. As a result, it might increase training efficiency to converge faster. 

\section{Conclusion and Future work}
In this study, we have developed LBKT, which employs a BERT-based architecture with an LSTM block for processing long-sequence data, and Rasch model-based embeddings for different difficulty levels of questions. Experiments show that LBKT outperforms baseline models on most benchmark datasets. We also conducted the speed performance experiment on the two large-scale datasets containing long-sequence data. The results suggest that LBKT could process long-sequence data faster and is more resource-efficient. Moreover, we conducted an ablation study for different components of LBKT. The results indicate that the LSTM component aided in improving the performance of dealing with long-sequence data. Furthermore, we conducted an analysis of the embedding strategy using t-SNE. The result shows that Rasch embedding could process the difficulty-level features effectively. In future work, we plan to improve the architecture to process more comprehensive data to satisfy the sustainable development requirement of increasing large-scale datasets containing long-sequence data emerging. Furthermore, we plan to build a more advanced embedding strategy to allow the model to process multiple data types simultaneously, such as exercises and concepts information from the dataset.
\section*{Acknowledgments}
This was was supported in part by......

%Bibliography
\bibliographystyle{unsrt}  
\bibliography{references}

\begin{thebibliography}{10}

\bibitem{abdelrahman2022knowledge}
Ghodai Abdelrahman, Qing Wang, and Bernardo~Pereira Nunes.
\newblock Knowledge tracing: A survey.
\newblock {\em ACM Computing Surveys}, 2022.

\bibitem{choi2020ednet}
Youngduck Choi, Youngnam Lee, Dongmin Shin, Junghyun Cho, Seoyon Park, Seewoo Lee, Jineon Baek, Chan Bae, Byungsoo Kim, and Jaewe Heo.
\newblock Ednet: A large-scale hierarchical dataset in education.
\newblock In {\em International Conference on Artificial Intelligence in Education}, pages 69--73. Springer, 2020.

\bibitem{chang2015modeling}
Haw-Shiuan Chang, Hwai-Jung Hsu, and Kuan-Ta Chen.
\newblock Modeling exercise relationships in e-learning: A unified approach.
\newblock In {\em EDM}, pages 532--535, 2015.

\bibitem{liu2021efficient}
Yang Liu, Jing Zhou, and Weiguo Lin.
\newblock Efficient attentive knowledge tracing for long-tail distributed records.
\newblock In {\em 2021 IEEE/ACIS 6th International Conference on Big Data, Cloud Computing, and Data Science (BCD)}, pages 104--109. IEEE, 2021.

\bibitem{jeon2021last}
SeungKee Jeon.
\newblock Last query transformer rnn for knowledge tracing.
\newblock {\em arXiv preprint arXiv:2102.05038}, 2021.

\bibitem{liu2021survey}
Qi~Liu, Shuanghong Shen, Zhenya Huang, Enhong Chen, and Yonghe Zheng.
\newblock A survey of knowledge tracing.
\newblock {\em arXiv preprint arXiv:2105.15106}, 2021.

\bibitem{ghosh2020context}
Aritra Ghosh, Neil Heffernan, and Andrew~S Lan.
\newblock Context-aware attentive knowledge tracing.
\newblock In {\em Proceedings of the 26th ACM SIGKDD international conference on knowledge discovery \& data mining}, pages 2330--2339, 2020.

\bibitem{devlin2018bert}
Jacob Devlin, Ming-Wei Chang, Kenton Lee, and Kristina Toutanova.
\newblock Bert: Pre-training of deep bidirectional transformers for language understanding.
\newblock {\em arXiv preprint arXiv:1810.04805}, 2018.

\bibitem{rasch1993probabilistic}
Georg Rasch.
\newblock {\em Probabilistic models for some intelligence and attainment tests.}
\newblock ERIC, 1993.

\bibitem{abdelrahman2023knowledge}
Ghodai Abdelrahman, Qing Wang, and Bernardo Nunes.
\newblock Knowledge tracing: A survey.
\newblock {\em ACM Computing Surveys}, 55(11):1--37, 2023.

\bibitem{corbett1994knowledge}
Albert~T Corbett and John~R Anderson.
\newblock Knowledge tracing: Modeling the acquisition of procedural knowledge.
\newblock {\em User modeling and user-adapted interaction}, 4(4):253--278, 1994.

\bibitem{corbett2000cognitive}
Albert Corbett.
\newblock Cognitive mastery learning in the act programming tutor.
\newblock In {\em Adaptive User Interfaces. AAAI SS-00-01. Retrieved from https://aaai. org/Library/Symposia/Spring/ss00-01. php}, 2000.

\bibitem{villano1992probabilistic}
Michael Villano.
\newblock Probabilistic student models: Bayesian belief networks and knowledge space theory.
\newblock In {\em International Conference on Intelligent Tutoring Systems}, pages 491--498. Springer, 1992.

\bibitem{yudelson2013individualized}
Michael~V Yudelson, Kenneth~R Koedinger, and Geoffrey~J Gordon.
\newblock Individualized bayesian knowledge tracing models.
\newblock In {\em International conference on artificial intelligence in education}, pages 171--180. Springer, 2013.

\bibitem{kaser2017dynamic}
Tanja K{\"a}ser, Severin Klingler, Alexander~G Schwing, and Markus Gross.
\newblock Dynamic bayesian networks for student modeling.
\newblock {\em IEEE Transactions on Learning Technologies}, 10(4):450--462, 2017.

\bibitem{cen2006learning}
Hao Cen, Kenneth Koedinger, and Brian Junker.
\newblock Learning factors analysis--a general method for cognitive model evaluation and improvement.
\newblock In {\em Intelligent Tutoring Systems: 8th International Conference, ITS 2006, Jhongli, Taiwan, June 26-30, 2006. Proceedings 8}, pages 164--175. Springer, 2006.

\bibitem{pavlik2009performance}
Philip~I Pavlik~Jr, Hao Cen, and Kenneth~R Koedinger.
\newblock Performance factors analysis--a new alternative to knowledge tracing.
\newblock {\em Online Submission}, 2009.

\bibitem{lecun2015deep}
Yann LeCun, Yoshua Bengio, and Geoffrey Hinton.
\newblock Deep learning.
\newblock {\em nature}, 521(7553):436--444, 2015.

\bibitem{piech2015deep}
Chris Piech, Jonathan Bassen, Jonathan Huang, Surya Ganguli, Mehran Sahami, Leonidas~J Guibas, and Jascha Sohl-Dickstein.
\newblock Deep knowledge tracing.
\newblock {\em Advances in neural information processing systems}, 28, 2015.

\bibitem{pandey2019self}
Shalini Pandey and George Karypis.
\newblock A self-attentive model for knowledge tracing.
\newblock {\em arXiv preprint arXiv:1907.06837}, 2019.

\bibitem{shin2021saint+}
Dongmin Shin, Yugeun Shim, Hangyeol Yu, Seewoo Lee, Byungsoo Kim, and Youngduck Choi.
\newblock Saint+: Integrating temporal features for ednet correctness prediction.
\newblock In {\em LAK21: 11th International Learning Analytics and Knowledge Conference}, pages 490--496, 2021.

\bibitem{nakagawa2019graph}
Hiromi Nakagawa, Yusuke Iwasawa, and Yutaka Matsuo.
\newblock Graph-based knowledge tracing: modeling student proficiency using graph neural network.
\newblock In {\em IEEE/WIC/ACM International Conference on Web Intelligence}, pages 156--163, 2019.

\bibitem{lee2022monacobert}
Unggi Lee, Yonghyun Park, Yujin Kim, Seongyune Choi, and Hyeoncheol Kim.
\newblock Monacobert: Monotonic attention based convbert for knowledge tracing.
\newblock {\em arXiv preprint arXiv:2208.12615}, 2022.

\bibitem{tiana2021bekt}
Zejie Tiana, Guangcong Zhengc, Brendan Flanaganb, Jiazhi Mic, and Hiroaki Ogatab.
\newblock Bekt: Deep knowledge tracing with bidirectional encoder representations from transformers.
\newblock In {\em Proceedings of the 29th International Conference on Computers in Education}, 2021.

\bibitem{kalyan2021ammus}
Katikapalli~Subramanyam Kalyan, Ajit Rajasekharan, and Sivanesan Sangeetha.
\newblock Ammus: A survey of transformer-based pretrained models in natural language processing.
\newblock {\em arXiv preprint arXiv:2108.05542}, 2021.

\bibitem{parmar2018image}
Niki Parmar, Ashish Vaswani, Jakob Uszkoreit, Lukasz Kaiser, Noam Shazeer, Alexander Ku, and Dustin Tran.
\newblock Image transformer.
\newblock In {\em International Conference on Machine Learning}, pages 4055--4064. PMLR, 2018.

\bibitem{floridi2020gpt}
Luciano Floridi and Massimo Chiriatti.
\newblock Gpt-3: Its nature, scope, limits, and consequences.
\newblock {\em Minds and Machines}, 30(4):681--694, 2020.

\bibitem{jiang2020convbert}
Zi-Hang Jiang, Weihao Yu, Daquan Zhou, Yunpeng Chen, Jiashi Feng, and Shuicheng Yan.
\newblock Convbert: Improving bert with span-based dynamic convolution.
\newblock {\em Advances in Neural Information Processing Systems}, 33:12837--12848, 2020.

\bibitem{sun2019bert4rec}
Fei Sun, Jun Liu, Jian Wu, Changhua Pei, Xiao Lin, Wenwu Ou, and Peng Jiang.
\newblock Bert4rec: Sequential recommendation with bidirectional encoder representations from transformer.
\newblock In {\em Proceedings of the 28th ACM international conference on information and knowledge management}, pages 1441--1450, 2019.

\bibitem{donahue2019lakhnes}
Chris Donahue, Huanru~Henry Mao, Yiting~Ethan Li, Garrison~W Cottrell, and Julian McAuley.
\newblock Lakhnes: Improving multi-instrumental music generation with cross-domain pre-training.
\newblock {\em arXiv preprint arXiv:1907.04868}, 2019.

\bibitem{li2023broader}
Zhaoxing Li, Mark Jacobsen, Lei Shi, Yunzhan Zhou, and Jindi Wang.
\newblock Broader and deeper: A multi-features with latent relations bert knowledge tracing model.
\newblock In {\em European Conference on Technology Enhanced Learning}, pages 183--197. Springer, 2023.

\bibitem{li2022simstu}
Zhaoxing Li, Lei Shi, Alexandra Cristea, Yunzhan Zhou, Chenghao Xiao, and Ziqi Pan.
\newblock Simstu-transformer: A transformer-based approach to simulating student behaviour.
\newblock In {\em International Conference on Artificial Intelligence in Education}, pages 348--351. Springer, 2022.

\bibitem{li2023towards}
Zhaoxing Li, Lei Shi, Yunzhan Zhou, and Jindi Wang.
\newblock Towards student behaviour simulation: a decision transformer based approach.
\newblock In {\em International Conference on Intelligent Tutoring Systems}, pages 553--562. Springer, 2023.

\bibitem{li2023deep}
Zhaoxing Li.
\newblock {\em Deep Reinforcement Learning Approaches for Technology Enhanced Learning}.
\newblock PhD thesis, Durham University, 2023.

\bibitem{li2023sim}
Zhaoxing Li, Lei Shi, Jindi Wang, Alexandra~I Cristea, and Yunzhan Zhou.
\newblock Sim-gail: A generative adversarial imitation learning approach of student modelling for intelligent tutoring systems.
\newblock {\em Neural Computing and Applications}, 35(34):24369--24388, 2023.

\bibitem{wang2023exploring}
Jindi Wang, Ioannis Ivrissimtzis, Zhaoxing Li, Yunzhan Zhou, and Lei Shi.
\newblock Exploring the potential of immersive virtual environments for learning american sign language.
\newblock In {\em European Conference on Technology Enhanced Learning}, pages 459--474. Springer, 2023.

\bibitem{wang2024comparative}
Jindi Wang, Ioannis Ivrissimtzis, Zhaoxing Li, and Lei Shi.
\newblock Comparative efficacy of 2d and 3d virtual reality games in american sign language learning.
\newblock In {\em The 31st IEEE Conference on Virtual Reality and 3D User Interfaces}. Newcastle University, 2024.

\bibitem{wang2023user}
Jindi Wang, Ioannis Ivrissimtzis, Zhaoxing Li, Yunzhan Zhou, and Lei Shi.
\newblock User-defined hand gesture interface to improve user experience of learning american sign language.
\newblock In {\em International Conference on Intelligent Tutoring Systems}, pages 479--490. Springer, 2023.

\bibitem{wang2024impact}
Jindi Wang, Ioannis Ivrissimtzis, Zhaoxing Li, and Lei Shi.
\newblock Impact of personalised ai chat assistant on mediated human-human textual conversations: Exploring female-male differences.
\newblock In {\em Companion Proceedings of the 29th International Conference on Intelligent User Interfaces}, pages 78--83, 2024.

\bibitem{tan2022bidkt}
Weicong Tan, Yuan Jin, Ming Liu, and He~Zhang.
\newblock Bidkt: Deep knowledge tracing with bert.
\newblock In {\em International Conference on Ad Hoc Networks, International Conference on Testbeds and Research Infrastructures}, pages 260--278. Springer, 2022.

\bibitem{sun2019muti}
Xia Sun, Xu~Zhao, Yuan Ma, Xinrui Yuan, Feijuan He, and Jun Feng.
\newblock Muti-behavior features based knowledge tracking using decision tree improved dkvmn.
\newblock In {\em Proceedings of the ACM Turing Celebration Conference-China}, pages 1--6, 2019.

\bibitem{zhang2021sequential}
Xuelong Zhang, Juntao Zhang, Nanzhou Lin, and Xiandi Yang.
\newblock Sequential self-attentive model for knowledge tracing.
\newblock In {\em International Conference on Artificial Neural Networks}, pages 318--330. Springer, 2021.

\end{thebibliography}

\end{document}